\documentclass[final]{custom}
\usepackage{graphicx}
\usepackage{rotating}
\usepackage{amssymb}
\usepackage{amsmath}
\usepackage{caption}
\usepackage{multicol}
\usepackage{float}
\usepackage{subcaption}
\usepackage{booktabs} % Top and bottom rules for table
\makeatletter

\begin{document}

\newcommand{\hdblarrow}{H\makebox[0.9ex][l]{$\downdownarrows$}-}
\title{Improvement of contact-less KID design using multilayered Al/Ti material for resonator.}

\author{%
    \name{J. Colas$^{1\star}$\thanks{$^{\star}$Email: j.colas@ip2i.in2p3.fr}, M. Calvo$^{2}$, J. Goupy$^{2}$, A. Monfardini$^{2}$, M. De Jesus$^{1}$, J. Billard$^{1}$, A. Juillard$^{1}$ L. Vagneron$^{1}$}%
    \affil{$^{1}$ \textit{Univ Lyon, Universit\'e Lyon 1, CNRS/IN2P3, IP2I-Lyon, F-69622, Villeurbanne, France}, \\ $^{2}$ \textit{Univ. Grenoble Alpes, CNRS, Grenoble INP, Institut N\'eel, 38000 Grenoble, France}}%
}

% \authorrunning{\maketitle}
%\titlerunning{\maketitle}
\maketitle

\begin{abstract}

    The necessity to increase exposure in rare event searches experiments by maintaining a low energy threshold and a good energy resolution leads to segmented detectors as in EDELWEISS (Dark Matter), CUORE (0$\nu\beta\beta$) or RICOCHET (CE$\nu$NS) for example. However, the large number of sub-elements can dramatically increase the complexity of such detector arrays. In this work we report on our progress towards designing a flexible detector technology based on KID resonators evaporated on massive target crystals readout by a contact-less feed-line. Providing that we achieve $\mathcal{O}(100)$~eV energy threshold, such approach could easily be scaled to tens of kilogram detector arrays thanks to the intrinsic multiplexing capability of mKIDs. Using a 30~g silicon target absorber with Al/Ti multilayers for the KID resonator, we report a significant improvement of our detector response exhibiting a keV-scale energy resolution combined with the absence of position dependence on the event location. Indeed, compared to our previous work, we are now able to properly identify calibration lines from surface (20~keV X-rays) and bulk events (60~keV gamma rays). This significant improvement  is an important step toward a better understanding of phonons and quasiparticles dynamics which is pivotal in optimizing this technology.

    \keywords{kinetic inductance detector, rare events searches}

\end{abstract}

\section{Motivations and experimental setup}

Multiplexing approaches are widely explored to overcome issues related to the scaling up of solid state cryogenic detectors. Although mKIDs, within the context of individual particle detection, are not at the same level of maturity as other sensing technologies, {\it e.g.} Ge-NTD or TES, they have the benefit of being intrinsically easy to multiplex. Therefore, this work focuses on improving our KID-based detector design energy resolution for future applications to low-energy and rare event searches.

%To give an example, these issues might be the cabling induced noise and heat flux inside the cryostat. Although one of the current best technology, in term of sensitivity and resolution, is the Transition Edge Sensor (TES) multiplexed using SQUID readout, the complexity of such system is still challenging and requires very special cares to reach state-of-the-art performances. In the other hand the Kinetic Inductance Detectors are comparatively very simple to build and operate, but the energy resolution is not as good as the best TES bolometers and more development are needed to reach better resolutions. 

%%%%%%%%%%%%%%%%%%%%%%%%%%%%%%%%%%%%%%%%%%%%%%%%%%%%%%%%%%%%%%%%%%%%%%%%%%%%%%%%%%%%%%%%%%%%%%%%%
%%%%%%%%%%%%%%%%%%%%%%%%%%%%%%%%%%%%%%%%%%%%%%%%%%%%%%%%%%%%%%%%%%%%%%%%%%%%%%%%%%%%%%%%%%%%%%%%%

\begin{figure}[htbp]
    \centering
    \includegraphics[width=0.8\linewidth, keepaspectratio]{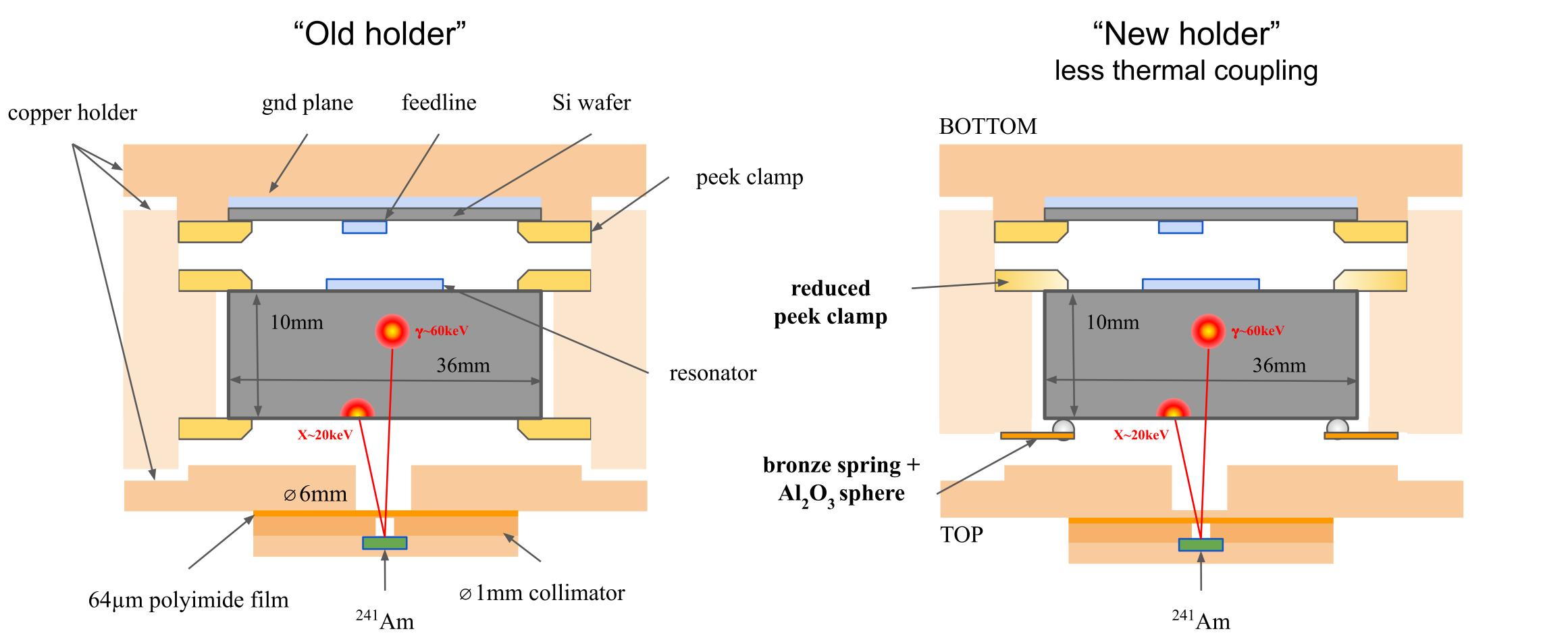}
    \caption{Schematic presentation of the two holder used in this study. The spacing between the resonator and the feedline is roughly 300$\mu$m. (Color figure online.)}
    \label{fig:holder}
\end{figure}

The proposed design discussed hereafter, called \textit{wifi-KID}, has a deported feed-line which is not on the same substrate as the resonator. The coupling between the feed-line and resonator is made through vacuum. The spacing between the resonator, already presented in our first wifi-KID study~\cite{Goupy2019}, and the feed-line is roughly 300~$\mu$m. The resonator is evaporated on the target material consisting of a silicon crystal of dimensions $36\times36\times10$~mm$^3$, and all of the parts are maintained inside a copper holder. We explored the effect of two strategies to hold the target crystal, which are both presented in Fig.~\ref{fig:holder}: the so-called ``old'' design based on peek clamps (left panel) and our ``new'' one (right panel) which uses springs and sapphire balls to reduce the thermal contact and possible phonon losses.

In our first wifi-KID work~\cite{Goupy2019} we used a pure 20~nm thick aluminium resonator, but the performance was not sufficient enough to distinguish the calibration sources properly. In the present work we report on the improvement in our detector response induced by the use of multilayered Al/Ti materials. We fabricated two new types of resonators: Ti-Al (10-25~nm) and Al-Ti-Al (15-30-30~nm). Note that the order of element follows the proximity of the target such that for the Ti-Al devices, the Ti is below the Al thin film.

\begin{figure}[htbp]
    \centering
    \includegraphics[width=0.7\linewidth, keepaspectratio]{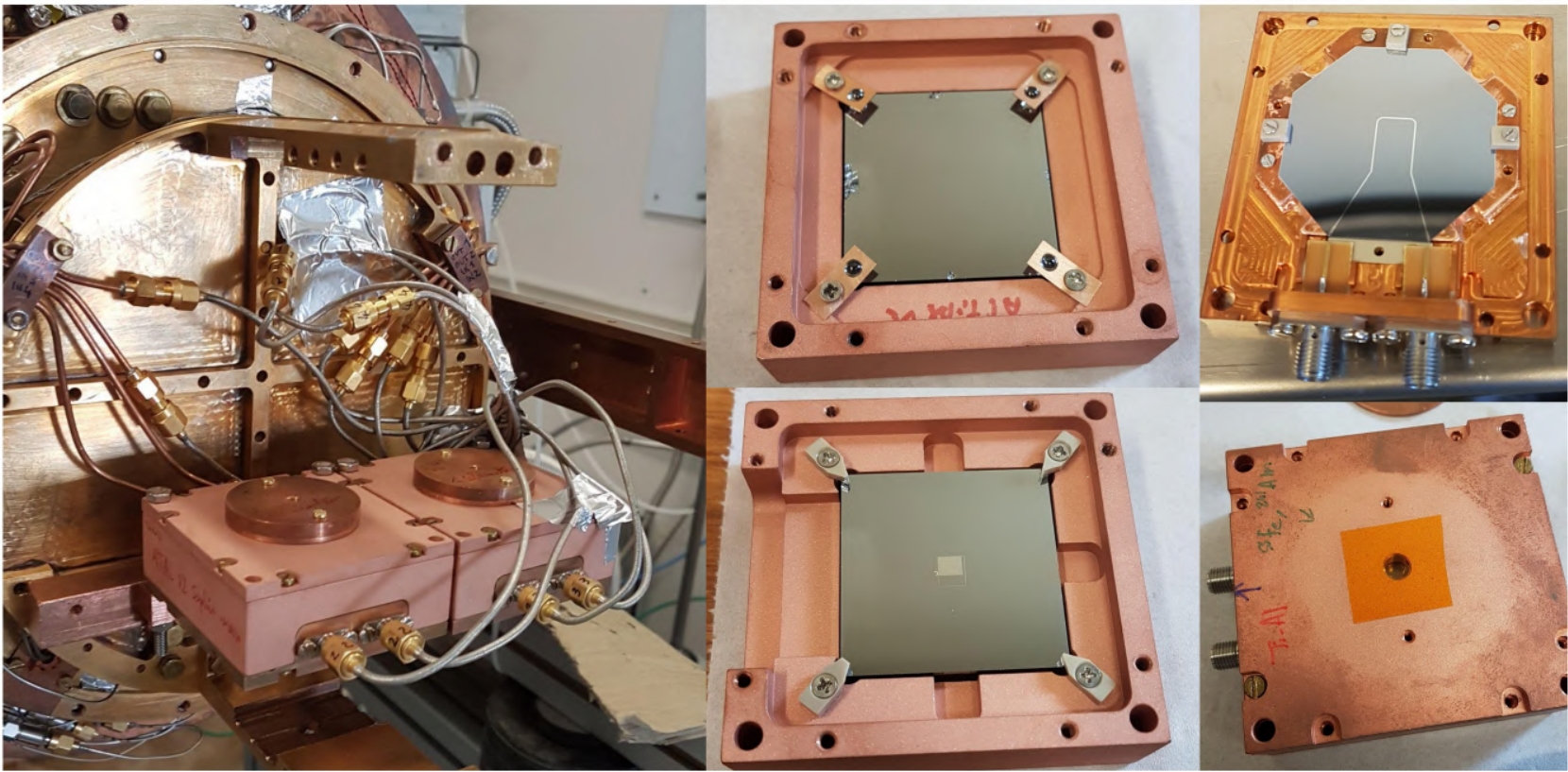}
    \caption{\textit{Left} - Two contactless KID detectors mounted inside the NIKA 1.5 cryostat. \textit{Right} - Four pictures  of different parts of the detector. (Color figure online.)}
    \label{fig:my_label}
\end{figure}
In the following, we present the resulting tests and characterisations of our four devices:
\begin{itemize}
    \item Al 20nm (first paper, with new analysis)
    \item Ti-Al 10-25nm (new holder)
    \item Al-Ti-Al 15-30-30nm (old holder)
    \item Al-Ti-Al 15-30-30nm (new holder)
\end{itemize}

Adding a layer of Ti for the resonator induces a lower $T_c$ and thus a lower superconducting energy gap. To achieve low enough temperatures we used another cryostat for multilayered resonator than the pure Al sensor. The base temperature for this dilution cryostat, which was used for NIKA 1.5 development \cite{nika1p5}, is around 90~mK and does not benefit from vibration decoupling system like low noise dry cryostat~\cite{Maisonobe_2018}.

The two calibration sources used was a $^{241}Am$ sample which produces alpha particles ($\sim$5~MeV), 60~keV gammas and low energy X-rays around 20~keV, and a $^{55}Fe$ radioactive sources producing $\sim$6~keV X-rays. Since the alpha particles are too energetic and lead to non-linear detector response we added a polyimide tape layer to stop them. In silicon the interaction length of 60~keV gamma is about 1~cm which is comparable to the target thickness, the gammas will deposit their energy uniformly in the target crystal. On the contrary, the emitted X-rays will interact only at the surface.

%%%%%%%%%%%%%%%%%%%%%%%%%%%%%%%%%%%%%%%%%%%%%%%%%%%%%%%%%%%%%%%%%%%%%%%%%%%%%%%%%%%%%%%%%%%%%%%%%
%%%%%%%%%%%%%%%%%%%%%%%%%%%%%%%%%%%%%%%%%%%%%%%%%%%%%%%%%%%%%%%%%%%%%%%%%%%%%%%%%%%%%%%%%%%%%%%%%

\section{Characterisation, calibration and performance}

We developed a dedicated characterisation and calibration procedure.
The first step is an equilibrium characterisation. This means that we study the characteristics of the resonator with respect to the thermal bath temperature of the cryostat. The idea is to evaluate the quality factors and the resonance frequency of the resonator at equilibrium for different temperature using a standard procedure~\cite{Probst2014}. As  explained in~\cite{Goupy2019}, measuring the relative detuning allows us to estimate the kinetic inductance ratio $\alpha$ as well as the energy gap $\Delta$.

%\begin{figure}[h!]
%    \centering
%    \includegraphics[width=0.7\linewidth, keepaspectratio]{images/resonance_circle_AlTiAl2.pdf}
%    \caption{\textit{Left} - IQ data in complex plane for multiple cryostat temperature. \textit{Right Top} - Transmission dependency with excitation frequency for multiple cryostat temperature. \textit{Right Bottom} - Phase dependency with excitation frequency for multiple cryostat temperature. (Color figure online.)}
%    \label{fig:circle}
%\end{figure}

\begin{figure}[h!]
    \centering
    \includegraphics[width=0.7\linewidth, keepaspectratio]{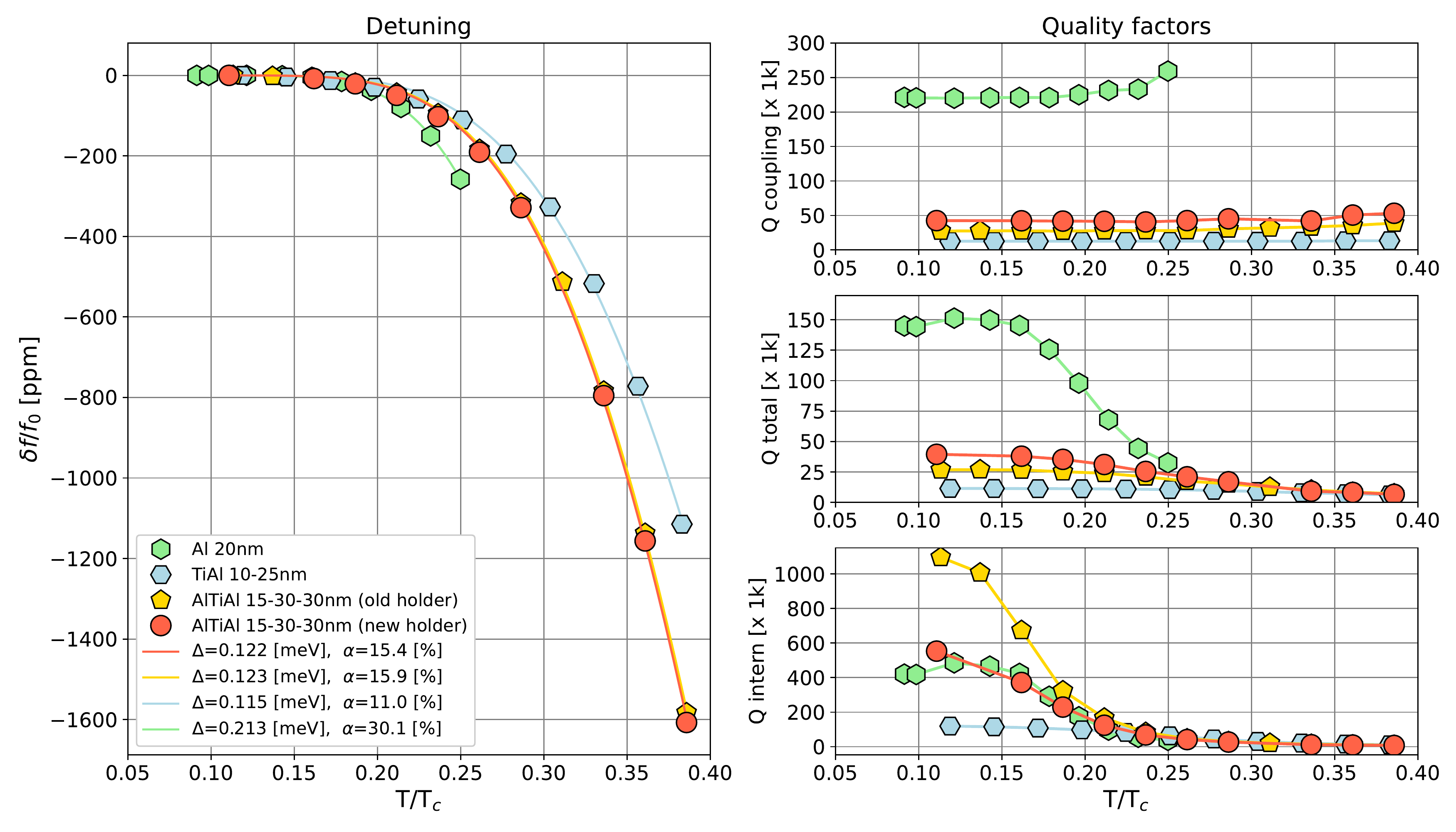}
    \caption{\textit{Left} - Relative detuning vs. reduced temperature for multiple devices. \textit{Right} - Quality factors for the corresponding devices, \textit{from top to bottom}: $\mathcal{Q}_\text{coupling}$, $\mathcal{Q}_\text{total}$ and $\mathcal{Q}_\text{internal}$. (Color figure online.)}
    \label{fig:Qfactors}
\end{figure}

The measurements performed on all of our devices are shown in Fig.~\ref{fig:Qfactors}. Because the critical temperature is material dependent we choose to use the reduced temperature $T/T_c$ to fairly compare all four devices.
%The $T_c$ plays a big role in the superconductor properties, in particular in the BCS theory where the energy gap is proportional to $T_c$~\cite{}. 
%What our measures say is that the pure aluminium device is the most sensitive, it’s expected since the sensitivity is inversely proportional to the thickness, and the Ti-Al device is the least sensitive. %The  estimated energy gaps and kinetic inductance ratio follow the expected behaviour as well :$\Delta_\text{Al} > \Delta_\text{Al-Ti} > \Delta_\text{Al-Ti-Al}$. 
Our results suggest that the pure aluminium device is the most sensitive, as one could have expected from its lower thickness, and that we successfully achieved high quality factors, especially for the Al-Ti-Al device with $Q_i \simeq 1e6$. Unfortunately, we found that the Ti-Al device did not work properly as it was exhibiting a very low overall quality factor and a degraded energy resolution compared to other devices. This device will therefore not be considered in the remaining of this work comparing the performance and response of our detector prototypes.

\begin{figure}[h!]
    \centering
    \includegraphics[width=\linewidth, keepaspectratio]{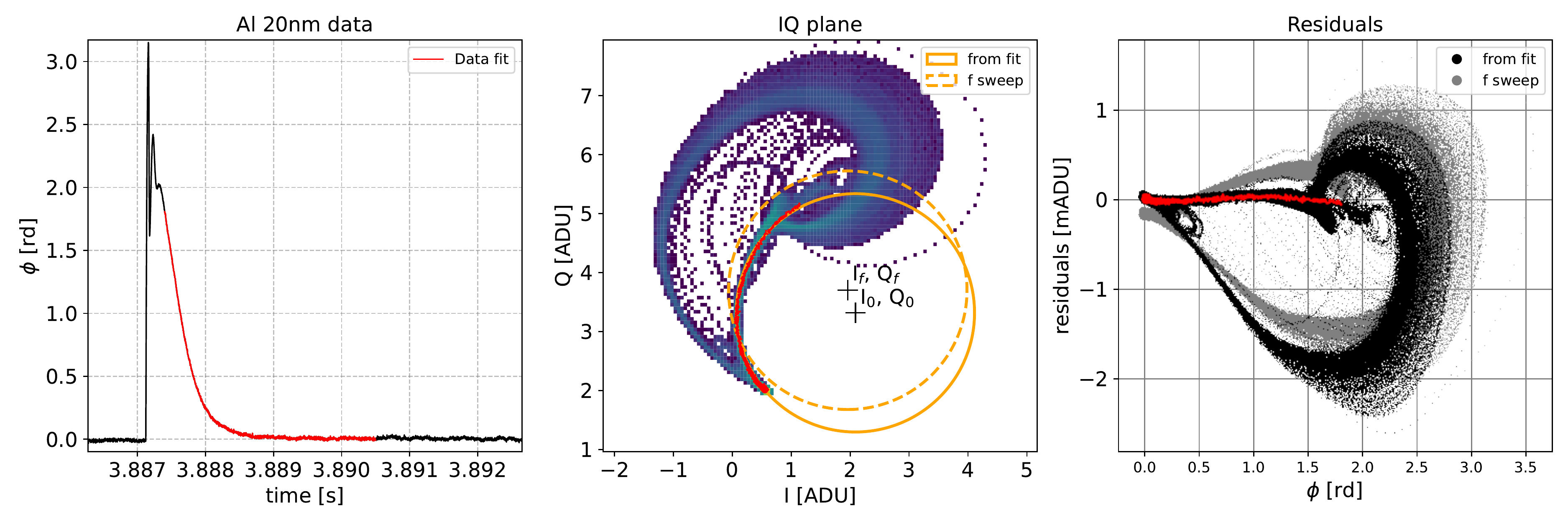}
    \caption{\textit{Left} - Zoom on temporal data window used for calibration, in \textit{red} the relaxation response used to calibrate. \textit{Middle} - Same representation as on the left but in the complex IQ plane with the optimal calibration circle in orange solid line centered in $I_c, Q_c$ and the frequency scan calibration circle in orange dashed line, centered in $I_f, Q_f$. \textit{Right} - Residuals of projected data on the calibration circle as a function of the phase shift $\theta$.}
    \label{fig:calib}
\end{figure}

This equilibrium characterisation allows us to gather a large number of information concerning our devices. However, the behaviour of resonators is slightly different for non thermal excitation and thus the calibration requires a dedicated out of equilibrium study, which is the second step of our characterisation phase. Usually the signal of one resonator is considered two dimensional, with a phase and an amplitude relatively to the resonant circle at the base temperature~\cite{calder}. The developed approach is to use the relaxation response of the resonator after a high energy deposition, {\it e.g.} a cosmic ray or a MeV-alpha particle interaction. We fitted this relaxation response in the complex plane with a circle discarding the non-linear part of its response. This procedure is illustrated in Fig.~\ref{fig:calib} where the temporal data are shown in the left panel, their representation in the complex plane in the central panel, and the residuals as a function of the phase shift $\theta$ from the fit in the right panel.

As can be derived from Fig.~\ref{fig:calib} the derived circle centered in ($I_0$,$Q_0$) fits accurately the data. This suggests that the bi-dimensional (phase and amplitude) detector response can be reduced to a single phase shift with respect to the ($I_0$,$Q_0$) circle to calibrate our data in the region of interest of small perturbations. This corresponding phase shift $\theta$ is defined as $\theta = \arctan((I-I_0)/(Q-Q_0))$. We found that is properly describes the a-thermal response of our sensor following energy depositions in the target crystal. It should be noted that in our current  experimental setup it is not possible to scan over a range of discrete energy depositions $E$ using optical fibre, we therefore only calibrated our devices using radioactive sources of $^{55}$Fe and $^{241}$Am. The measurement of $\theta(E)$ is hence only accessible for two energies, which is not sufficient to fully characterise the linearity of detector energy-scale.

\begin{figure}[h!]
    \centering
    \includegraphics[width=0.45\linewidth, keepaspectratio]{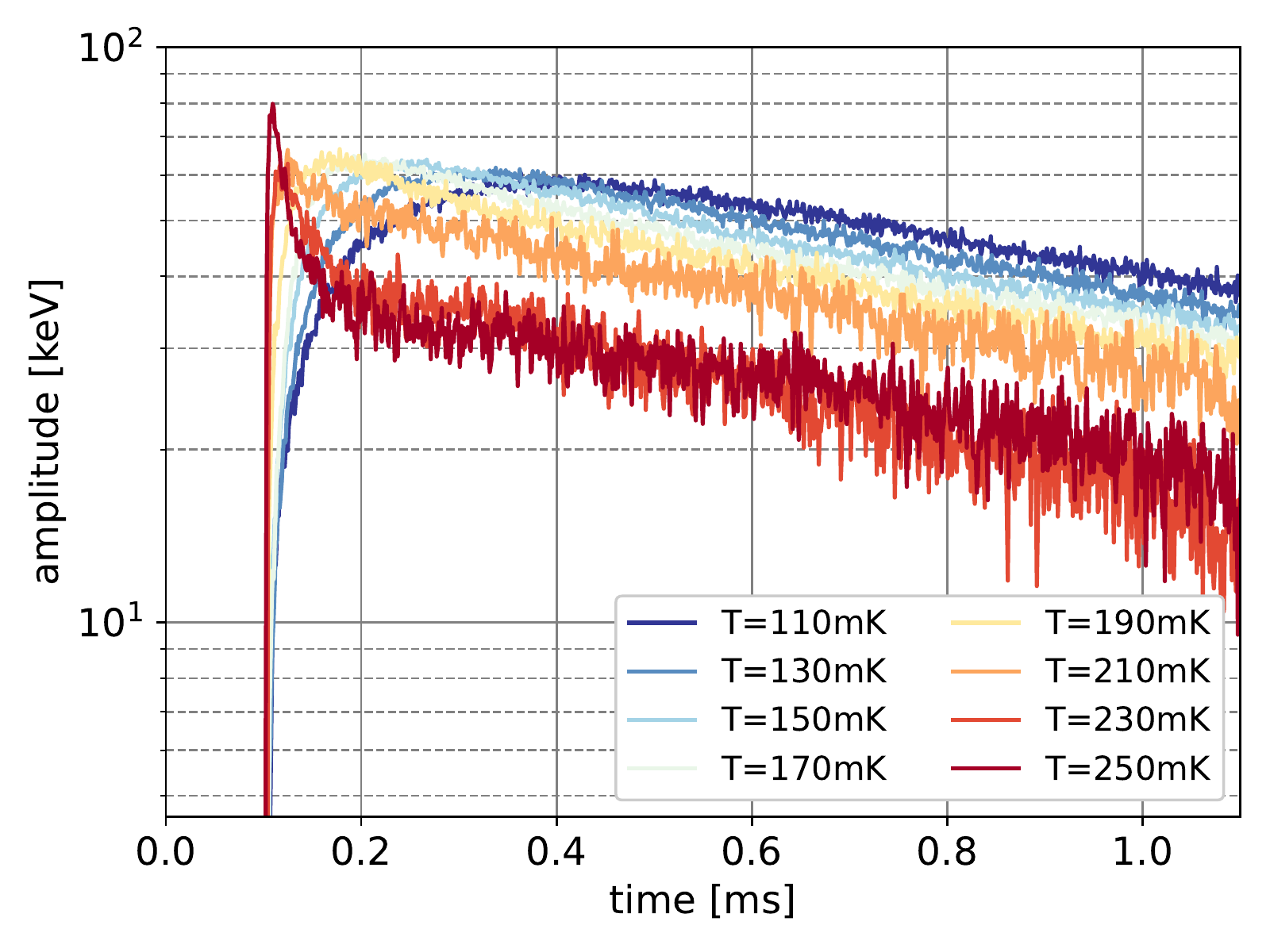}
    %includegraphics[width=0.48\linewidth, keepaspectratio]{images/pulses_shape_revised.pdf}
    \includegraphics[width=0.44\linewidth, keepaspectratio]{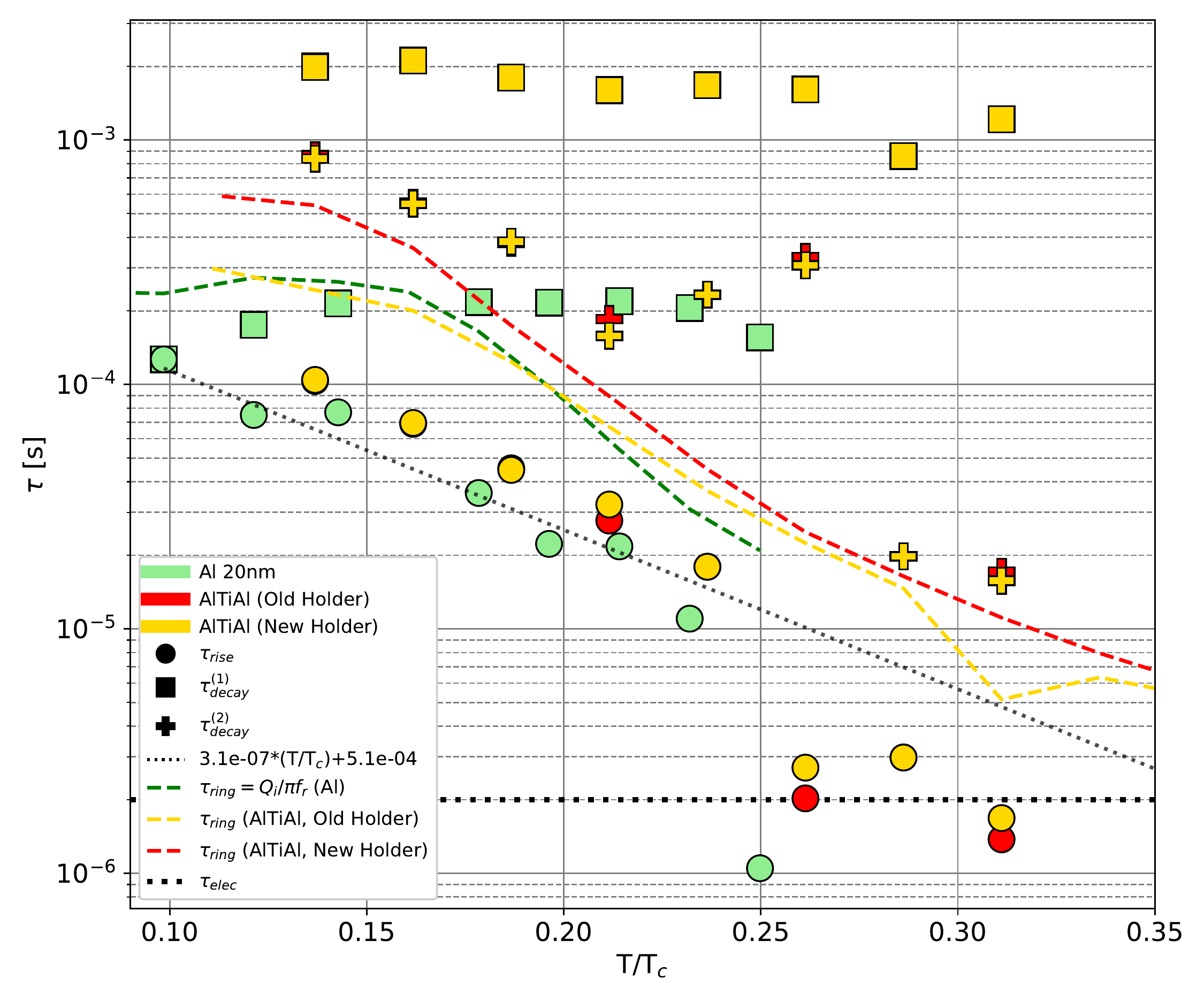}
    \caption{\textit{Left} - Evolution of mean pulse with temperature, \textit{y-axis in log scale}. \textit{Right} - Characteristic times fitted for a model of three exponential $\tau_\text{rise}$ (\textit{circle}), $\tau_\text{decay}^{2}$ (\textit{cross}) $\tau_\text{decay}^{1}$ (\textit{square}) for multiple detectors at different reduced temperature $T/T_c$ : Al 20nm (\textit{green}), AlTiAl 15-30-30nm (\textit{yellow, red}). (Color figure online.)}
    \label{fig:pulse_shape}
\end{figure}

Measuring $\theta$ (derived from the $I$ and $Q$ observables) continuously over time and performing offline signal processing, including offline triggering, allowed us to extract characteristic pulse shapes for each bath temperatures. The data processing pipeline used to estimate the amplitude of the pulses is based on optimal filtering and was primarily developed for the Ricochet experiment~\cite{mps}.
%Regarding the amplitude spectrum obtained using the Al 20nm resonator we were convinced that there is a sort of position dependency of the measured amplitude which widen the calibration peaks and destroy our ability to reconstruct deposited energies. The use of multilayered Al-Ti films allowed us to change the dynamic of our detector, as shown in Fig. . 
Figure~\ref{fig:pulse_shape} shows the evolution of the three characteristic time constants (one rise, and two decays) resulting from our pulse fitting procedure for various temperatures and all three detector configurations: one pure Al 20~nm and two Al-Ti-Al 15-30-30~nm in two different holders. Our fitting model is motivated by the three expected response time constants from the quasi-particle recombination rate, the resonator ring time and the phonon lifetime inside the target. We see that at the lowest relative temperature, only two characteristic time constants (one rising and one decaying) are required to fit the data. However, at the highest relative temperatures, a third characteristic time appears to be required. Interestingly, we see that the longest decay time constant appears to be independent on the bath temperature, while both the rise and the second decay time constants decrease significantly with the relative temperature, suggesting that the former is related to the phonon absorption rate. As a matter of fact, the difference in the $\tau_{\text{decay}}^{(1)}$ between the Al and Al-Ti-Al devices can be explained by considering the lower gap for Al-Ti-Al resonator implying more phonons to break Cooper pairs and slowing down the pulse relaxation. The rise time seems related to the resonator ring time which is expected to decrease with the increase of temperature (see the dotted lines). Finally the interpretation of the third characteristic time remains uncertain as it appears difficult with the available data to break the degeneracy between the quasiparticle lifetime and the resonator ring time. Further studies dedicated to better understand the origin of our observed time constants are ongoing.

%This shows that there is a competition between multiple processes governed by three characteristic times : the quasi-particle recombination time, the resonator ring time and the phonons lifetime. The identification of each of these processes remain uncertain but our current best hypothesis is that the decay time is limited by the phonons lifetime because there is no temperature dependency. The one order of magnitude increase between Al and Al-Ti-Al resonator may be explained by the lower energy gap for Al-Ti-Al which enhance the collection of phonons. The rise time, on the other hand, seems to follow the same evolution with temperature normalised by the superconductivity critical temperature and may be linked to the resonator ring time. At higher temperature the ring time become small enough to make the recombination of quasi-particle visible. The possible explanation for this is to consider that the recombination of quasi-particle can induce phonons emission with energy close to the energy gap an this can leads to a phonons excess with a decay following the quasi-particle recombination time. %\textbf{Est-ce qu'il faut faire un fit de cette composante ? }

%For high temperature a simple model with two characteristic times is not suited to fit accurately the pulses. 

Eventually, slowing down the resonator decay time by reducing the energy gap greatly improved the phonon collection of our detector simultaneously reducing the position dependence on the event location in the crystal. As shown in Fig.~\ref{fig:spectrum}, thanks to the use of Al/Ti/Al multilayers, we can properly recover both surface and bulk calibration lines from low-energy X-rays and 60~keV gammas, respectively. Figure~\ref{fig:spectrum} also shows the comparison between our observed energy spectrum (red) and the Geant4 simulations (black) smeared with our observed energy resolution with an additional energy dependent correction factor of 1.2\% to correct for detector non-linearity at increasing energies.

%  \begin{figure}[H]
%    \includegraphics[width=0.5\linewidth, keepaspectratio]{images/data_vs_simu_N1P5139.pdf}
%    \caption{Comparison between Geant4 simulation (\textit{black}) and data obtained with a AlTiAl (15-30-30nm) (\textit{red}) resonator with two calibration sources, $^{55}$Fe and $^{241}$Am. (Color figure online.)}
%    \label{fig:spectrum}
%  \end{figure}

%\begin{center}
%\begin{tabular}{l l l l l}
%\toprule
%\textbf{Material} & \textbf{Al} & \textbf{Ti-Al} & \textbf{Al-Ti-Al} \\
%\midrule
%\textbf{Thick. [nm]} & 20 &  10-25 & 15-30-30 \\
%\textbf{$T_c$ [K]} & 1.40 & 0.76 & 0.80 \\
%\textbf{$\Delta$ [$\mu$eV]} & 213 & 115 & 123 \\
%\textbf{$\alpha$ [\%]} & 30 & 11 & 15.5 \\
%\textbf{$f_0$ [MHz]} & 564.57 & 577.18 & 589.36 \\
%\textbf{$\sigma$ [keV]} & ~2 & \textbf{N.A.} & ~1.5 \\
%\bottomrule
%\end{tabular}
%\captionof{table}{ Summary table of the state of the art of wifi-KID project.}
%\label{table:sum}
%\end{center}

\begin{table}
    \begin{minipage}{0.53\linewidth}
        \includegraphics[width=\linewidth, keepaspectratio]{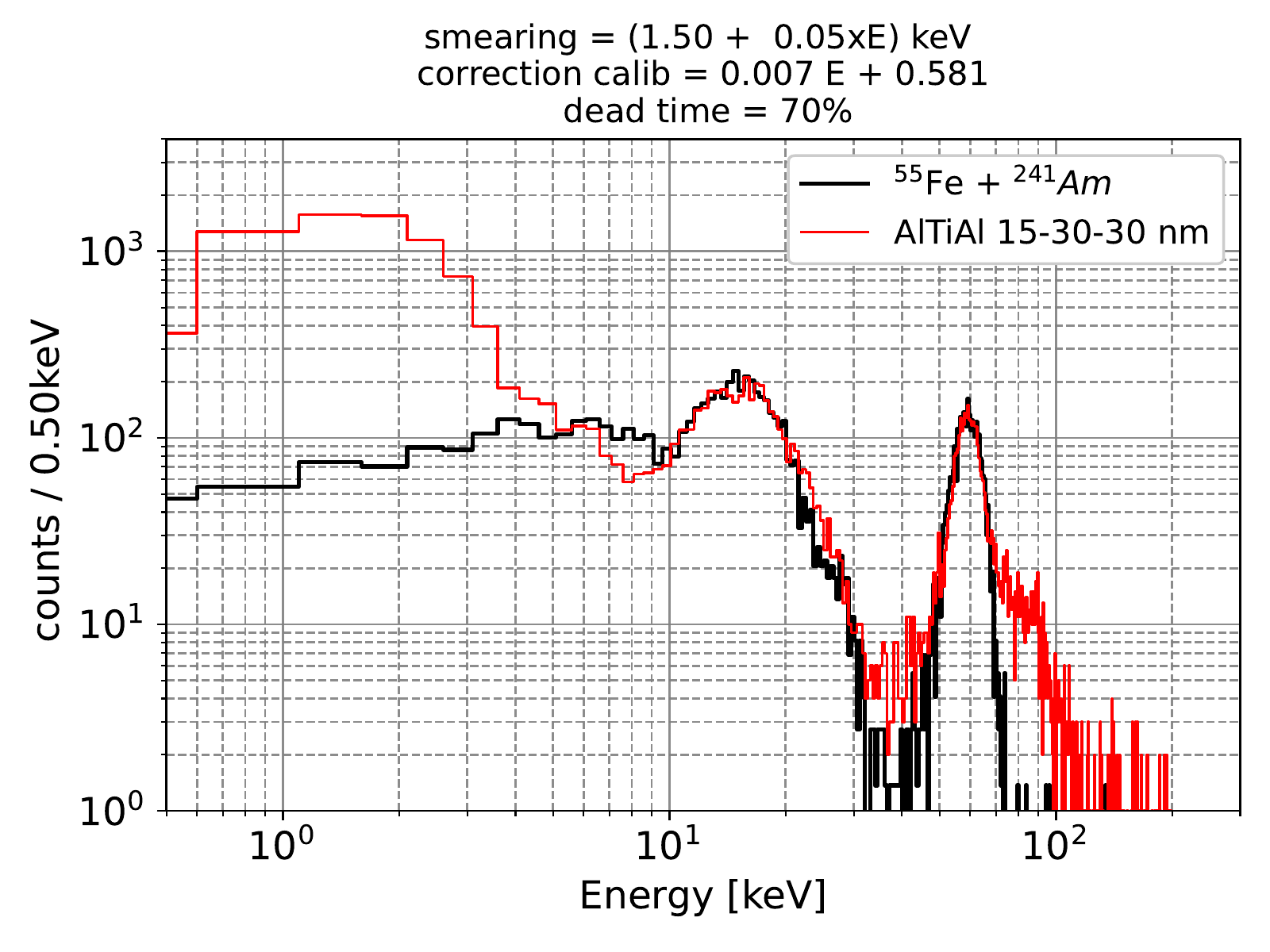}
        \captionof{figure}{Comparison between Geant4 simulation (\textit{black}) and data obtained with a AlTiAl (15-30-30nm) (\textit{red}) resonator with two calibration sources, $^{55}$Fe and $^{241}$Am. (Color figure online.)}
        \label{fig:spectrum}
    \end{minipage}\hfill
    \begin{minipage}{0.40\linewidth}
        \centering
        \resizebox{\textwidth}{!}{%
            \begin{tabular}{l l l l l}
                \toprule
                \textbf{Material}           & \textbf{Al} & \textbf{Ti-Al} & \textbf{Al-Ti-Al} \\
                \midrule
                \textbf{Thick. [nm]}        & 20          & 10-25          & 15-30-30          \\
                \textbf{$T_c$ [K]}          & 1.40        & 0.76           & 0.80              \\
                \textbf{$\Delta$ [$\mu$eV]} & 213         & 115            & 123               \\
                \textbf{$\alpha$ [\%]}      & 30          & 11             & 15.5              \\
                \textbf{$f_0$ [MHz]}        & 564.57      & 577.18         & 589.36            \\
                \textbf{$\sigma$ [keV]}     & ~2          & \textbf{N.A.}  & ~1.5              \\
                \bottomrule
            \end{tabular}}
        \captionof{table}{Summary of the various results obtained by the wifi-KID investigators.}
        \label{table:sum}
    \end{minipage}
\end{table}

\vspace{-.5cm}
\section{Conclusion}

The use of multilayered Al/Ti material improved the performances of the wifi-KID in terms of position dependency. We are now able to identify the calibration peaks in the amplitude spectrum, which was not possible with the previous design in pure aluminium~\cite{Goupy2019}. With a reported 1.5~keV baseline resolution, these devices are still not sufficiently good to claim that our prototypes are competitive in the field of low-energy and rare event searches. But this work is a significant step towards the optimisation of this cryogenic detector technology with proven and reliable multiplexing capability for future highly segmented large detector arrays.

\section*{Acknowledgements}
This work is supported by the European Research Council (ERC) under the European Union’s Horizon 2020 research and innovation program under Grant Agreement ERC-StG-CENNS 803079

\pagebreak

% \bibliographystyle{plain} % We choose the &quot;plain&quot; reference style
% \bibliography{bib}

\end{document}